\documentclass[aps,twocolumn,prb,showpacs,10pt,floatfix,groupedaddress]{revtex4-1}
\usepackage{amssymb}
\usepackage{amsmath}
\usepackage{graphicx}
\usepackage{dcolumn}
\usepackage{bm}
\usepackage{textcomp}

\begin{document}

\title{Anisotropic AC conductivity of strained graphene}

\author{M. Oliva-Leyva}
\email{moliva@fisica.unam.mx}
\author{Gerardo G. Naumis}
\email{naumis@fisica.unam.mx}

\affiliation{Departamento de F\'{i}sica-Qu\'{i}mica, Instituto de
F\'{i}sica, Universidad Nacional Aut\'{o}noma de M\'{e}xico (UNAM),
Apartado Postal 20-364, 01000 M\'{e}xico, Distrito Federal,
M\'{e}xico}


\begin{abstract}

The density of states and the AC conductivity of graphene
under uniform strain are calculated using a new Dirac hamiltonian
that takes into account the main three ingredients that change the
electronic properties of strained graphene: the real displacement of
the Fermi energy, the reciprocal lattice strain and the changes in
the overlap of atomic orbitals. Our simple analytical expressions of
the density of states and the AC conductivity generalizes previous
expressions only available for uniaxial strain. The results suggest
a way to measure the Gr\"{u}neisen parameter $\beta$ that appears in
any calculation of strained graphene, as well as the emergence of a
sort of Hall effect due to shear strain.

\end{abstract}


\maketitle


Graphene,\cite{Novoselov04} a two-dimensional form of carbon, has
attracted an enormous attention of both experimental and theoretical
studies to understand and take advantage of its remarkable
properties.\cite{Geim09,Novoselov11} Among its most exotic
properties, one can cite a linear band dispersion of charge carriers
at the so-called Dirac points, hence in graphene the charge carries
behavior as massless Dirac fermions.\cite{Neto09} This has been
suggested the possibility of studying phenomena originally predicted
for relativistic particle physics in such a unique condensed-matter
system.

Its mechanical properties, strength and flexibility, are also
unique. Graphene is the second strongest material ever measured (the
first is Carbyne), with an effective Young's modulus of $\sim 1$TPa,
and can reversibly withstand elastic deformations up to $25\%$.\cite{Lee08}
This unusual interval of elastic response results in a
peculiar interplay between its electronic and morphological
properties, which has opened a new opportunity to explore the strain-induced
modifications of the electronic properties of graphene: strain
engineering.\cite{Pereira09b,Guinea12,Zhan12} This concept has a
successful history in the strained silicon technology.

In the literature, the most popular approach to study the electronic
implications of graphene lattice deformations is based on a
combination of tight-binding (TB) model of the electrons and linear
elasticity
theory.\cite{Pereira09a,Cocco10,Choi10,Kitt13,Barraza13a,Barraza13b,deJuan12,deJuan13,Oliva-Leyva13,Gomes12,Yan13}
Among the strain-induced implications most investigated is the
opening of a gap at the Fermi level because of its importance for
the functionality of graphene-based logic devices. For uniaxial
strain, being the case usually considered, it has been shown that
the strain-induced opening of a band gap depends critically on the
direction of strain and requires values as large as
$23\%$.\cite{Pereira09a} Nevertheless, it has been shown that a
combination of shear and uniaxial strain can be used to open a gap
for more easily accessible reversible deformations.\cite{Cocco10}
Needless to say, other strain implication is the anisotropy in the
electrons dynamics, which has been theoretical and experimentally
established.\cite{deJuan12,deJuan13,Oliva-Leyva13,Gomes12,Yan13}
Most recently, within the Dirac-cone approximation and considering
the shift of the Dirac points, an expression was given to the
anisotropic Fermi velocity as a function of the strain
tensor.\cite{Oliva-Leyva13} This new result enables us to generalize
the alternating-current AC conductivity expression for strained
graphene, and thereby, to propose possible applications. The AC
conductivity is one of the most useful tools to obtain a deeper
insight into the graphene electronic properties, and thus, to infer
details of their morphology.

Numerous theoretical works and several experiments have been devoted
to investigate the optical conductivity properties of
graphene.\cite{Gusynin06,Gusynin07a,Ziegler06,Ziegler07,Stauber08,Pellegrino10,Pereira10,Naumis11,Horng11,Gornyi12,Herbut13,Scharf13,
Tan07,Mak08,Li09,Mak12} Gusynin \emph{et al.}\cite{Gusynin06}
presented analytical expressions for the AC conductivity in
unstrained graphene. Pellegrino \emph{et al.}\cite{Pellegrino10}
reported the uniaxial strain effects on the AC conductivity within
the tight-binding approximation. Pereira \emph{et
al.}\cite{Pereira10} also studied the uniaxial strain effects on the
AC conductivity, but within the Dirac-cone approximation. However,
in the graphene literature does not exit an available AC
conductivity expression of graphene under an uniform strain, which
can be straightforwardly used for comparison with experimental data.
This issue is the principal motivation for our study.

The second motivation is that recently, it has been found that the usual approach
of using pseudomagnetic fields to treat strain contains several problems, since
the Fermi energy falls far from the place where the usual approximations
are made. A new Dirac Hamiltonian has been proposed
to solve this problem taking into account the main three
ingredients that change the
electronic properties of strained graphene: the real displacement of
the Fermi energy, the reciprocal lattice strain and the changes in
the overlap of atomic orbitals.\cite{Oliva-Leyva13} In previous studies of the subject,
the first ingredient was missing,\cite{Oliva-Leyva13} while the second one only very
recently has been discovered. As a result, many of the results
concerning experimental situations need to be revised. Here we
provide the first application of this new Hamiltonian, since
we are able to give an experimental proposal to measure the Gr\"{u}neisen
parameter $\beta$ which is a vital quantity to perform calculations
in strained graphene.

In this letter, first we present the Dirac-like model used for
uniformly strained graphene. Then we study the density of states of
strained graphene, which is required in our posterior calculation of
the AC conductivity by means of the Kubo formula. Finally, the
conclusions are given.

\begin{figure}[h,t]
\includegraphics[width=8cm]{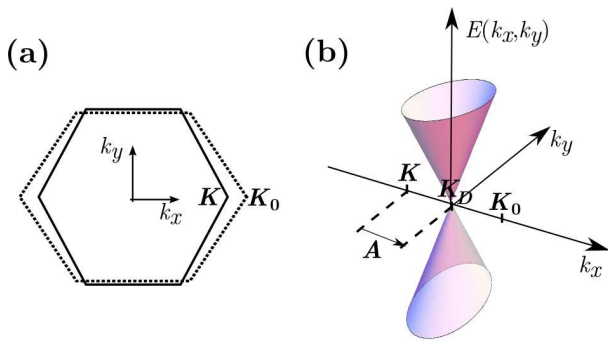}
\caption{\label{fig1}(Color online) (a) $\bm{K}_{0}$ ($\bm{K}$) is a
high-symmetry point of the first Brillouin zone, denoted by the
dashed (solid) line, for unstrained (strained) graphene. The
strained Brillouin zone correspond to stretched graphene in the $x$
direction. (b) Sketch of the strain-induced shift of the
$\bm{K}_{D}$ Dirac point and the distorted Dirac cone. The
displacement of $\bm{K}_{D}$ from $\bm{K}$ is given by the vector
$\bm{A}$.}
\end{figure}

\emph{Strained graphene model used}: From a combination of a tight-binding description and linear
elasticity theory, recently it has been reported that the low-energy
Dirac Hamiltonian for electrons in graphene under uniform strain is
given by\cite{Oliva-Leyva13}
\begin{equation}\label{H}
H\simeq \hbar
v_{0}\bm{\sigma}\cdot(\bar{\bm{I}}-\beta\bar{\bm{\epsilon}}+\bar{\bm{\epsilon}})\cdot\bm{q},
\end{equation}
where $v_{0}$ is the Fermi velocity for the undeformed lattice;
$\bm{\sigma}=(\sigma_{x},\sigma_{y})$, the two Pauli matrices;
$\bar{\bm{I}}$, the $2\times2$ identity matrix;
$\bar{\bm{\epsilon}}$, the strain tensor; $\beta$, the Gr\"{u}neisen
parameter; and $\bm{q}$, the momentum measured relative to the
$\bm{K}_{D}$ Dirac point,\cite{Oliva-Leyva13} which does not
coincide with the $\bm{K}$ high-symmetry points of the strained
Brillouin zone, as detailed in FIG.~\ref{fig1}. To first order in
strain the shift of the $\bm{K}_{D}$ Dirac point can
be written as,\cite{Oliva-Leyva13}
\begin{equation}
\bm{K}_{D}\simeq\bm{K}+\bm{A}\simeq(\bar{\bm{I}}+\bar{\bm{\epsilon}})^{-1}\cdot\bm{K}_{0}+\bm{A},
\end{equation}
where $\bm{K}_{0}$ is a high-symmetry point of the unstrained
Brillouin zone and $\bm{A}$ is related to the strain tensor
$\bar{\bm{\epsilon}}$ by
\begin{equation}\label{VP}
A_{x}=\frac{\beta}{2a_{0}}(\bar{\epsilon}_{xx}-\bar{\epsilon}_{yy}),
\ \ \ A_{y}=-\frac{\beta}{2a_{0}}(2\bar{\epsilon}_{xy}),
\end{equation}
with $a_{0}$ being the unstrained carbon-carbon distance. Note that,
here the vector $\bm{A}$ is related to the Dirac cone shift, whereas in
the theory of the strain-induced pseudomagnetic field it is
interpreted as a pseudovector potential.\cite{Neto09}

Let us emphasize the three contributions appearing in (\ref{H}):
$\hbar v_{0}\bm{\sigma}\cdot\bm{q}$ is the low-energy Dirac
Hamiltonian for unstrained graphene; the $\beta$-dependent term
$-\hbar v_{0}\beta\bm{\sigma}\cdot\bar{\bm{\epsilon}}\cdot\bm{q}$ is
associated to the strain-induced changes in hopping parameters; and
the $\beta$-independent term $\hbar
v_{0}\bm{\sigma}\cdot\bar{\bm{\epsilon}}\cdot\bm{q}$ is only a
consequence of the reciprocal space distortion due to lattice
deformation.

From Eq.~(\ref{H}) one can recognize a direction-dependent Fermi
velocity, which can be written in tensorial notation as
\begin{equation}\label{V}
\bar{\bm{v}}=v_{0}(\bar{\bm{I}}-\beta\bar{\bm{\epsilon}}+\bar{\bm{\epsilon}}).
\end{equation}

For testing Eq.~(\ref{V}), let us consider the seemingly trivial
case of an isotropic strain
$\bar{\bm{\epsilon}}_{i}=\epsilon\bar{\bm{I}}$. Under
$\bar{\bm{\epsilon}}_{i}$, the new carbon-carbon distance is
$a=a_{0}(1+\epsilon)$. On the other hand, the new hopping parameter
to first order in strain is $t=t_{0}(1-\beta\epsilon)$. Therefore,
the new Fermi velocity is
\begin{equation*}
v=\frac{3ta}{2\hbar}\simeq
\frac{3t_{0}a_{0}}{2\hbar}(1-\beta\epsilon+\epsilon)\simeq
v_{0}(1-\beta\epsilon+\epsilon),
\end{equation*}
and in tensorial notation,
$\bar{\bm{v}}=v_{0}(\bar{\bm{I}}-\beta\bar{\bm{\epsilon}}_{i}+\bar{\bm{\epsilon}}_{i})$,
which can directly obtained from the generalized expression
(\ref{V}).

Since the Fermi velocity, the most important parameter in graphene
physics, is changed by the strain, one would expect changes in the conductivity.


Another ingredient that influences the conductivity is the density
of states (DOS). Furthermore, the DOS is accessible through experiments of
scanning tunneling microscopy, and thus, can serve to test the new
Dirac Hamiltonian, Eq.~(\ref{H}).

The DOS for graphene under uniform strain can be computed in
momentum space as the trace of the Green's function
\begin{equation}\label{DOS}
\rho(E)=\int\int \text{Tr} [E-H(q_{x},q_{y})]^{-1}\text{d} q_{x}\text{d} q_{y},
\end{equation}
with $H(q_{x},q_{y})$ being the Hamiltonian (\ref{H}). This
computation is straightforward if one proposes the change of
integration variables
\begin{equation}\label{CV}
\bm{q}=(\bar{\bm{I}}-\beta\bar{\bm{\epsilon}}+\bar{\bm{\epsilon}}_{i})^{-1}\cdot\bm{q}^{*}.
\end{equation}
This transformation yields,
\begin{equation}\label{DOST}
\rho(E)=\int J \text{Tr} [E-H_{0}(q_{x}^{*},q_{y}^{*})]^{-1}\text{d}
q_{x}^{*}\text{d} q_{y}^{*},
\end{equation}
with $J$ being the Jacobian determinant of Eq.~(\ref{CV}) and
$H_{0}(q_{x}^{*},q_{y}^{*})=\hbar v_{0}\bm{\sigma}\cdot\bm{q}^{*}$ being
the unstrained Hamiltonian. It can be proved that the Jacobian
determinant is given as $J=1/\det(\bar{\bm{v}}/v_{0})$. 
Then, for the DOS immediately it follows
\begin{equation}
\rho(E)=\rho_{0}(E)v_{0}^{2}\det(\bar{\bm{v}}/v_{0}),
\end{equation}
where $\rho_{0}(E)$ is the DOS for unstrained graphene and reads
$\rho_{0}(E)=2|E|/(\pi \hbar^{2}v_{0}^{2})$. Finally, up to first
order to in strain the DOS gives
\begin{equation}\label{DOSF}
\rho(E)\simeq\rho_{0}(E)(1+\beta\text{Tr}\bar{\bm{\epsilon}}-\text{Tr}\bar{\bm{\epsilon}}),
\end{equation}
which is a generalized expression available for any uniform strain,
and not only for uniaxial strain, as done in previous
works.\cite{Pereira10}

Since $\beta-1>0$, the strain effect in the DOS dependents on the
sign of $\text{Tr}\bar{\bm{\epsilon}}$. Note that, up to first order in strain
$\text{Tr}\bar{\bm{\epsilon}}$ can be written as
$\text{Tr}\bar{\bm{\epsilon}}=A/A_{0}-1$, $A$ being the area of strained
graphene sample and $A_{0}$, the unstrained area sample. Thus, for
an expanded sample ($A/A_{0}>1$) the strain effect in the DOS is a
slope enhancement, whereas for ($A/A_{0}<1$) the effect is a slope
worsen. It is also worth stressing that for the case of shear
strain, i.e. $\bar{\epsilon}_{xx}=\bar{\epsilon}_{yy}=0$ and
$\bar{\epsilon}_{xy}=\bar{\epsilon}_{yx}=\epsilon$, the DOS does not
change.


Let us now obtain the AC conductivity $\bar{\bm{\sigma}}(w)$ of
graphene under uniform strain by combining Eq.~(\ref{H}) and the
Kubo formula, assuming a linear response to a external electric
field with frequency $w$. Following the representation used in
Refs.~[22,23], the frequency-dependent conductivity tensor
$\bar{\bm{\sigma}}(w)$ can be written as a double integral with
respect to two energies $E$, $E'$:
\begin{eqnarray} \label{C}
\bar{\sigma}_{ij}(w)=\frac{i}{\hbar}\int\int \text{Tr}\{ j_{i}\delta(H-E')j_{j} \delta(H-E)\} \nonumber\\
\qquad\times\frac{1}{E-E'+w-i\alpha}\frac{f(E)-f(E')}{E-E'}\text{d}
E\text{d} E',
\end{eqnarray}
where $f(E)=(1+\exp[E/(k_{B}T)])^{-1}$ is the Fermi function at
temperature $T$ and $j_{l}=-i e [H,r_{l}]$ is the current operator
in the $l$-direction, with $l=x,y$.

To calculate the integral (\ref{C}) is convenient once again to use the change of
integration variables given by Eq.~(\ref{CV}). In the new variables
$(q_{x}^{*},q_{y}^{*})$, the Hamiltonian transforms as
$H=H_{0}(q_{x}^{*},q_{y}^{*})$ and the current operator components,
as
\begin{eqnarray}
j_{x}&=&-i e[H,r_{x}]=e\frac{\partial H}{\partial q_{x}}, \nonumber\\
&=&e\left(\frac{\partial H}{\partial q_{x}^{*}} \frac{\partial
q_{x}^{*}}{\partial q_{x}} + \frac{\partial H}{\partial q_{y}^{*}}
\frac{\partial q_{y}^{*}}{\partial q_{x}}\right), \nonumber\\
&=&(1-\tilde{\beta}\epsilon_{xx})j_{x}^{*}
-\tilde{\beta}\epsilon_{xy}j_{y}^{*},
\end{eqnarray}
and
\begin{equation}
j_{y}=(1-\tilde{\beta}\epsilon_{yy})j_{y}^{*}
-\tilde{\beta}\epsilon_{xy}j_{x}^{*},
\end{equation}
where $j_{x}^{*}=e(\partial H/\partial q_{x}^{*})$ and
$j_{y}^{*}=e(\partial H/\partial q_{y}^{*})$ are the current
operator components for unstrained graphene and
$\tilde{\beta}=\beta-1$. Plugging these expressions into
Eq.~(\ref{C}) and calculating up to first order in strain, we obtain
\begin{eqnarray}
\bar{\sigma}_{xx}(w)&\simeq&(1-2\tilde{\beta}\bar{\epsilon}_{xx})J\sigma_{0}(w),\label{xx}\\
\bar{\sigma}_{yy}(w)&\simeq&(1-2\tilde{\beta}\bar{\epsilon}_{yy})J\sigma_{0}(w),\\
\bar{\sigma}_{xy}(w)&=&\bar{\sigma}_{yx}(w)\simeq-2\tilde{\beta}\bar{\epsilon}_{xy}J\sigma_{0}(w),
\end{eqnarray}
with
\begin{eqnarray} \label{C0}
\sigma_{0}(w)=\frac{i}{\hbar}\int\int \text{Tr}\{ j_{i}^{*}\delta(H_{0}-E')j_{i}^{*} \delta(H_{0}-E)\} \nonumber\\
\qquad\times\frac{1}{E-E'+w-i\alpha}\frac{f(E)-f(E')}{E-E'}\text{d}
E\text{d} E'.
\end{eqnarray}
Note that, the frequency-dependent conductivity tensor of unstrained
graphene $\bar{\bm{\sigma}}_{0}(w)$ is given by
$\bar{\bm{\sigma}}_{0}(w)=\sigma_{0}(w)\bar{\bm{I}}$, as has been
calculated in Refs. [20--23]. Finally, from
Eqs.~(\ref{xx})--(\ref{C0}) it follows that the frequency-dependent
the conductivity tensor of graphene under uniform strain can be
written as
\begin{equation}\label{CVF}
\bar{\bm{\sigma}}(w)\simeq\sigma_{0}(w)(\bar{\bm{I}}-2\tilde{\beta}\bar{\bm{\epsilon}} + \tilde{\beta}\text{Tr}(\bar{\bm{\epsilon}})\bar{\bm{I}}).
\end{equation}

Our expression (\ref{CVF}) generalizes the reported results in
Ref.~[26] for case of a uniaxial strain, and allows us to make a
quick evaluation of the AC conductivity of graphene under any
uniform strain configuration. However, due to the approximations
considered one may wonder: How dependable are our linear corrections
in Eq.~(\ref{CVF})? Pereira \emph{et al.}\cite{Pereira10} showed
from \emph{ab initio} calculation for uniaxially strained graphene
that $\bar{\bm{\sigma}}(w)$ linearly dependents on strain up to at
least a $10\%$, which is a broad range of stretching where the
linear corrections are quite dependable.

\begin{figure}[h,t]
\includegraphics[width=8cm]{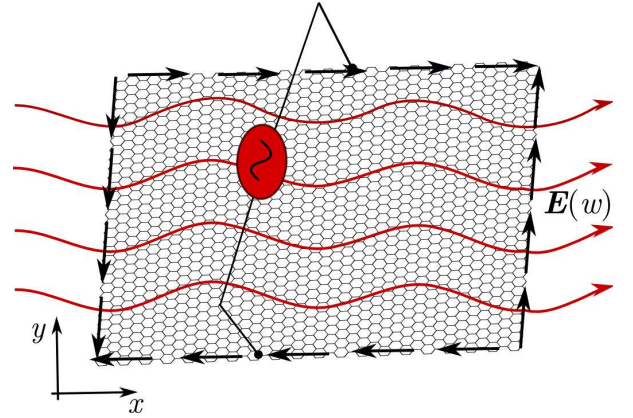}
\caption{\label{fig2}(Color online) Sketch of the observation of
Hall voltage in a rectangular graphene sample under shear strain and
a oscillating electric field $\bm{E}(w)$ in the $x$ direction
(curved red lines). Black arrows at sample boundaries represent the
forces field that leads to the shear deformation. The Hall voltage
is measured in the $y$ direction.}
\end{figure}

Unlike the DOS, the AC conductivity is modified for shear strain.
From Eq.~(\ref{CVF}), a sort of Hall effect is predicted due to
shear strain as sketched in FIG.~\ref{fig2}, since a component
$\sigma_{xy}(w)$ appears whenever $\epsilon_{xy}\neq0$. This result
is novel because strain-induced Hall effect in graphene has only
been discussed in nonuniform strain configurations.

Also, a possible application follows from Eq.~(\ref{CVF}). If the
conductivity can be measured with precision for different strain
configurations one can obtain by means of Eq.~(\ref{CVF}) a experimental
value of the Gr\"{u}neisen parameter $\beta$. For example, if the
conductivity $\sigma_{xx}$ of a graphene sample is measured for
uniaxial strains in the $x$ direction of $1\%$, $2\%$, $3\%$,...;
the scope of the lineal graph of $\sigma_{xx}$ vs strain is
$-(\beta-1)$. The knowledge of $\beta$ is vital for the
characterization of the strain-induced effects on the electronic
band structure.

In conclusion, using the Kubo formula and a new Dirac Hamiltonian
that takes into account the real strain-induced displacement of the
Fermi energy, we have derived the AC conductivity for graphene under
uniform strain. As a token, the corresponding expression gives a way
to measure the Gr\"{u}neisen parameter $\beta$ that appears in any
calculation of strained graphene. Also, we have reported a
generalized expression of the density of states, which does not
change for shear strain. However, under shear strain we reported a
sort of Hall effect which had not been discussed for uniform strain.

We thank the DGAPA-UNAM project IN-102513. M. Oliva-Leyva
acknowledges a scholarship from CONACyT (Mexico).


\bibliographystyle{unsrt}
\bibliography{GrapheneBiblioN}

\end{document}